\documentclass[prd, superscriptaddress, nofootinbib]{revtex4}
\usepackage{graphicx, epsfig}
\begin{document}

\def\be{\begin{equation}}
\def\ee{\end{equation}}
\def\ba{\begin{eqnarray}}
\def\ea{\end{eqnarray}}
\def\bq{\begin{quote}}
\def\eq{\end{quote}}
\def\PL{{ \it Phys. Lett.} }
\def\PRL{{\it Phys. Rev. Lett.} }
\def\NP{{\it Nucl. Phys.} }
\def\PR{{\it Phys. Rev.} }
\def\MPL{{\it Mod. Phys. Lett.} }
\def\IJMP{{\it Int. J. Mod .Phys.} }
\newcommand{\labell}[1]{\label{#1}\qquad_{#1}} 
\newcommand{\labels}[1]{\vskip-2ex$_{#1}$\label{#1}} 
\newcommand\gapp{\mathrel{\raise.3ex\hbox{$>$}\mkern-14mu
\lower0.6ex\hbox{$\sim$}}}
\newcommand\gsim{\gapp}
\newcommand\gtsim{\gapp}
\newcommand\lapp{\mathrel{\raise.3ex\hbox{$<$}\mkern-14mu
\lower0.6ex\hbox{$\sim$}}}
\newcommand\lsim{\lapp}
\newcommand\ltsim{\lapp}
\newcommand\M{{\cal M}}
\newcommand\order{{\cal O}}

\newcommand\extra{{\rm {extra}}}
\newcommand\FRW{{\rm {FRW}}}
\newcommand\brm{{\rm {b}}}
\newcommand\ord{{\rm {ord}}}
\newcommand\Pl{{\rm {pl}}}
\newcommand\Mpl{M_{\rm {pl}}}
\newcommand\mgap{m_{\rm {gap}}}
\newcommand\gB{g^{\left(\rm \small B\right)}}

\newcommand\Flow{F_{low}}
\newcommand\low{\rm low}
\newcommand\changed{\bf}

\title{Using Supernovae to Determine the Equation of State of the Dark Energy:
\\ Is Shallow Better Than Deep?}
\author{David N. Spergel}
\affiliation{Princeton University Observatory\\ Princeton NJ 08544}
\author{Glenn D. Starkman}
\affiliation{Case Western Reserve University \\
Cleveland, OH 44106-7079, USA}


\begin{abstract}
Measurements of the flux and redshifts of Type Ia supernovae have 
provided persuasive evidence that the expansion of the universe 
is accelerating.  If true, then in the context of standard FRW cosmology
this suggests that the energy density of the universe is dominated by
``dark energy" -- a component with negative pressure of magnitude 
comparable to its energy density.  To further investigate this 
phenomenon, more extensive surveys of supernovae are being planned.
Given the likely  timescales for completion, by the time 
data from these surveys are available some important cosmological parameters
 will be known to high precision from CMB measurements.
Here we consider the impact of that foreknowledge on the 
design of supernova surveys.  In particular we show that, 
despite greater opportunities to multiplex, purely from the 
point of view of statistical errors, 
a deep survey may not obviously be better than a shallow one.

\end{abstract}
\pacs{}

\rightline{CWRU-P5-02}

\maketitle



\section{Introduction}
Over the past few years, measurements of the flux and redshifts of 
Type Ia  supernova have provided increasingly persuasive evidence that
the expansion of the universe is accelerating \cite{SNCP, HRSS}.
The natural inference -- that the energy density of the universe is 
dominated  by some form of vacuum energy  -- has important consequences
for  both cosmology and particle physics.  Of particular interest is
the exact nature of the ``dark energy.'' Is it the
so-called cosmological constant, the energy density of the true vacuum
state of the universe? Is it predominantly the potential energy density 
of a new form of energy often called quintessence? or is it something
else entirely?   Since these each have dramatic, and dramatically different, implications
for fundamental physics  there is a 
clear need both to raise the confidence level of the conclusion
that the expansion is accelerating, and to better characterize the
time history of that expansion so that the nature of the 
``dark energy,'' in particular its equation of state, can be better understood.

\section{CMB As a Probe of Cosmological Parameters}

The recent expansion history of the universe (after matter-radiation equality)  --
the evolution of the scale factor $a$ with time $t$, or with conformal time
$\eta$ ($d t = a(\eta) d\eta$) -- is given by the Friedman equation:
\be
\label{Friedman}
\frac{da}{d\eta}\!=\!\sqrt{\Omega_o\!H_o^{\!2}\!}
\sqrt{\frac{a}{a_o}\!+\!\frac{\Omega_{c}}{\Omega_o}\!\left(\!\frac{a}{a_o}\!\right)^{\!2\!}\!+\!
\frac{1\!-\!\Omega_o\!-\!\Omega_{c}}{\Omega_o}\!\left(\!\frac{a}{a_o}\!\right)^{\!(\!1\!-\!3\!w\!)\!}}
.
\ee
The evolution of the expansion  is therefore characterized by $\Omega_o$, the current  ratio
of the matter energy density to the critical energy density,
$\Omega_{c}$, the current  ratio
of the curvature ``energy density'' to the critical energy density,
$H_o$, the current value of the Hubble parameter, and $w$ the equation of state parameter
for the dark energy. ($p_{dark} = w \rho_{dark}$.)

Great success has already been achieved in determining some of these parameters by a number
of techniques, and even greater precision is likely in the near future. 
We will focus in particular on measurements of  temperature anisotropies
in the cosmic microwave background radiation (CMB).
The positions of the acoustic peaks in the angular power spectrum of the CMB
depend on $\eta_{ls}$ (the conformal lookback time to the last-scattering surface)
and on the geometry of the universe, parametrized by $\Omega_{c}$.  
The shapes of  the peaks depend sensitively on $\Omega_o h_o^2$.
Current data on the first three peaks \cite{Boomerang,MAXIMA,DASI}
already argue strongly for a flat (or nearly flat) universe 
($\Omega_{\rm curvature}\simeq 0$).  
With anticipated data from the MAP satellite (and elsewhere)
the shape of the spectrum should also be well measured.
Thus the positions and shapes of the first peaks
in the CMB spectrum will soon be used to confine models of the dark energy
to a one-dimensional surface parametrized by $w$ \cite{parameters}. 
Extracting $w$ from CMBR measurements is far more difficult --
the value of $w$ has too little effect on the power spectrum of the anisotropies.

\section{Supernovae as Standard Candles}

While the absolute luminosity of supernovae is not known, properly 
calibrated Type Ia supernovae seem to be excellent standard candles 
\cite{HRSS,SNCP}.  
Measurements of the flux and redshift from a 
distant supernova when compared to measurements of similar 
supernovae at low redshift
($z_{\low}\ll1$)  measures the square of the ratio of the luminosity distances 
$d_L = \eta (1+z)$ as a function  of $z$:
\be
\label{FbyFlow}
\frac{F(z,w)}{F(z_{\low},w)} = \left[\frac{\eta(z) (1+ z)}{\eta(z_{\low}) (1+z_{\low})} \right]^{-2}
\ee
It has already been possible, by measuring $\eta(z)$ from the ground for a reasonable sample of 
supernovae, to argue convincingly that $w<0$ for a significant fraction
of the energy density of the universe \cite{HRSS,SNCP}.  

\section{Optimum Survey Strategies for Measuring $w$}

Accepting the desirability of improving the determination of $w$
(and perhaps more generally $w(z)$),  and working with the assumption
that Type Ia supernovae remain the best existing standard candles,
one can still ask what is the optimum possible survey strategy?
The answer depends very much on the precise goals of the survey \cite{previous}.
One option is to seek independent confirmation of the pre-existing 
determinations of $\Omega H_o^2$ and $\Omega_c$ while simultaneously
measuring $w$ and possibly its derivatives.  Alternatively, one could
accept the pre-existing measurements of $\Omega H_o^2$ and $\Omega_c$
and confine one's time and attention to the best possible measurement
of $w$ itself.  Admittedly, there are clear benefits to independent 
measurements of cosmological parameters. Nevertheless, we would argue 
that the uncertainty in the theoretical models of supernovae, coupled 
with likely observational uncertainties  make future supernova surveys
less powerful as opportunities for independent checks on existing and 
near-term CMB-based  determinations of parameters than they are as 
opportunities to break parameter degeneracies that the CMBR
measurements cannot or cannot easily  break and thus determine quantities
like $w$.  

The implications of this choice of survey strategies may be significant.
From a purely statistical viewpoint,
a supernova survey designed to measure
$\Omega H_o^2$, $\Omega_c$ {\bf and} $w$ will necessarily extend to higher z,
because it must measure the expansion history over a wide range of $z$ 
in order to make independent determinations of all three parameters
characterizing the evolution of the scale factor. However, 
except to address issues of systematic effects,
a survey designed to measure $w$ only, may not need to extend
to such a high redshift, since it seeks to measure only one parameter.
We shall  therefore pursue this option further.

Assuming perfect knowledge of the CMBR, we can fix the value of $ \Omega_o H_0^2$ and $\Omega_c$,
and  of the function $\Omega(w)$.  The former two are obtained from a direct fit
to the CMB angular power spectrum, while $\Omega(w)$ is obtained as follows.
For the purposes of illustration, we shall take $\Omega_c=0$.
The conformal distance  to the surface of last scatter is then
\be
\eta_{ls} \simeq \frac{1}{\sqrt{\Omega_oH_o^2}}\int_0^1 \frac{d x}{\sqrt{x + \frac{1-\Omega_o}{\Omega_o}x^{1-3w}}}  \ ,
\ee
(where  we have  taken $a_{ls}\simeq0$).
Fixing $\eta_{ls}$, this can be solved numerically for $\Omega_o(w)$; however, a 
linear fit can be obtained in the neighborhood of any $w$ as follows.
Differentiating with respect to $\Omega_o$, holding $\Omega_oH_o^2$ and $w$ fixed,
\be
\frac{\partial \eta_{ls}}{ \partial\Omega_o}\Bigr\vert_{\Omega_oH_o^2, w}
\simeq \frac{1}{\sqrt{\Omega_oH_o^2}}\frac{1}{2\Omega_o^2} 
\int_0^1 \frac{x^{1-3w} d x}{\left[x + \frac{1-\Omega_o}{\Omega_o}x^{1-3w}\right]^{3/2}}  \ .
\ee
Similarly,
differentiating with respect to $w$, holding $\Omega_oH_o^2$ and $\Omega_o$ fixed,
\be
\frac{\partial \eta_{ls}}{ \partial w}\Bigr\vert_{\Omega_oH_o^2, w}
\simeq \frac{1}{\sqrt{\Omega_oH_o^2}}\frac{3 (1-\Omega_o) }{2\Omega_o} 
\int_0^1 \frac{x^{1-3w} \ln(x) d x}{\left[x + \frac{1-\Omega_o}{\Omega_o}x^{1-3w}\right]^{3/2}}  \ .
\ee
Evaluating the ratio of these quantities at, for example,  $\Omega_o=0.3$, $w=-1$, we find
\be
\Omega_o(w) \simeq 0.3 + 0.274 (w + 1 ) \ .
\ee
Now, given measurements of the flux ratio $F(z,w)/F(z_{\low},w)$ 
between distant  supernovae and a sample of low redshift supernova, 
we can determine $w$.
The uncertainty in the determination of $w$ from one measurement of $F(z)/F(z_{\low})$ is
\be
\sigma_w 
= \left[\frac{d \ln (F/\Flow)}{d w}\right]^{-1}_{\Omega_oH_o^2,\Omega(w)} 
\frac{\sigma_{(F/\Flow)}}{F/\Flow} 
\ee
Using the relationship  (\ref{FbyFlow}), 
and assuming that any variance is dominated by the uncertainty in the flux determination
(i.e.  the uncertainty in the redshift is negligible), 
we can relate $\sigma_w$ to $\eta$.
For each measurement
\be
\sigma_w 
= \frac{1}{2}
\frac{\sigma_{(F/\Flow)}/\left(F/\Flow\right)}{\left\vert\frac{\partial\ln\eta}{\partial w}
- \frac{\partial\ln\eta_{\low}}{\partial w}\right\vert} .
\ee
For a survey with a fixed observing time, $\sigma_w$ is reduced by the square root of the number of supernovae $N_{SN}$
which can be probed, so that
\be
\sigma_w^{(survey)} = 
\frac{\sigma_{(F/\Flow)}/\left(F/\Flow\right)}{\left\vert\frac{\partial\ln\eta}{\partial w}
- \frac{\partial\ln\eta_{\low}}{\partial w}\right\vert} .
\frac{1}{\sqrt{N_{SN}}} \ .
\ee

Here we begin to see the impact of alternative observing strategies.
For a simple  survey with a simple instrument, the number of supernovae which one can 
observe at a given redshift will be proportional simply to the ratio of the flux of a supernova at the
redshift to a fixed threshold  flux -- $N_{SN} \propto F/F_{thresh}$.  However, it is possible
to improve on this strategy, especially at high redshifts, by increasing the field of view of the survey instrument,
monitoring more than one galaxy at a time for supernovae, and possibly following up with spectra in a fully multi-plexed
program. In this case the number of supernova discovered will scale as 
\be 
N_{SN} \propto F ~ \eta^2 \Delta \eta \propto (1 + z)^{-2} \ .
\ee
We have taken $\Delta \eta$ -- the range in conformal depth of the survey -- to be a constant.  
Other assumptions are possible. While they change the detailed dependence of the statistical errors 
on redshift, they do not change the qualitative behavior.

For a fully multiplexed survey
\be
\sigma_w^{(survey)} \propto \frac{\sigma_{(F/\Flow)}/(F/\Flow)}
	{\left\vert \frac{\partial \ln \eta}{\partial w}  -
	\frac{\partial \ln \eta_{\low}}{\partial w}\right\vert_{\Omega_oH_o^2,\Omega(w)}     }
	(1+z)
	\ .
\ee
Now, in the neighborhood of $w=-1$,
\be
\eta\vert_{\Omega_oH_o^2}\simeq \frac{1}{\sqrt{\Omega_oH_o^2}}
\int_{1/(1+z)}^1 \frac{d x}{ \sqrt{x + \frac{7}{3} x^4}}
\ee
and 
\begin{eqnarray}
\left(\frac{d \eta}{d w}\right)_{\Omega_oH_o^2}  
\simeq \frac{1}{2\sqrt{\Omega_oH_o^2}} \times & \cr
\times \int_{1/(1+z)}^1 \frac{x^4 d x}{ \left(x + \frac{7}{3} x^4\right)^{3/2}}&\left( 7 \ln(x) + \frac{0.274}{(0.3)^2}\right) \ .
\end{eqnarray}
These integrals can be evaluated numerically without difficulty,
allowing us to calculate $\sigma_w^{(survey)}$ as a function of the redshift of the
observations for a variety of $w$'s (cf. Fig.~1).
{\bf 
The principal point of interest is that while $\sigma_w^{(survey)}$ has a minimum at 
$z\simeq 1$ (somewhat lower for $w<-1$, somewhat higher for $w>-1$), 
this minimum is exceedingly shallow, so that the value of $\sigma_w$ at
$z=0.3$ is less than twice the minimum value. Moreover, our calculations
involved the supernova discovery rate.  However, it is in fact the follow up  spectroscopy
which is more demanding of telescope time; this scales as a higher power of  redshift 
(approximately $(1+z)^6$ according to some experts).  This shifts
the minimum down further in $z$. }
These results are relatively insensitive to 
the value of $w$, and to reasonable values of $z_{\low}$.

We have thus shown that, from a statistical point of  view, 
there is not the strong preference one might have suspected to measure
$w$ at high redshift --
intermediate redshifts ($z=0.3-0.5$) are  nearly as good for discovering
supernovae, and  at least as good for carefully studying them.
The length of time that one must integrate on individual distant supernovae 
offsets  the benefits one might get from the greater individual utility of 
these supernova and the larger accessible number of such supernova.
The details of the survey statistics will no doubt change as
one designs a more realistic survey, but the  broad conclusion will not -- for measuring $w$
with supernova, statistics are as good or better at low $z$ than at high $z$.  
The low z surveys offer the advantage of smaller K corrections and
the ability to study the host galaxy properties in detail with a wide
array of ground based telescopes.  For the low z surveys to be competitive
with the high z surveys, they require even more exquisite control of
systematic photometric errors. Ultimately, detailed trade studies are
needed to determine the best strategy for determining the properties of 
the dark energy. These studies should include our growing knowledge
about the geometry and composition of the  universe in their analyses.

\begin{figure}
\hbox{\psfig{figure=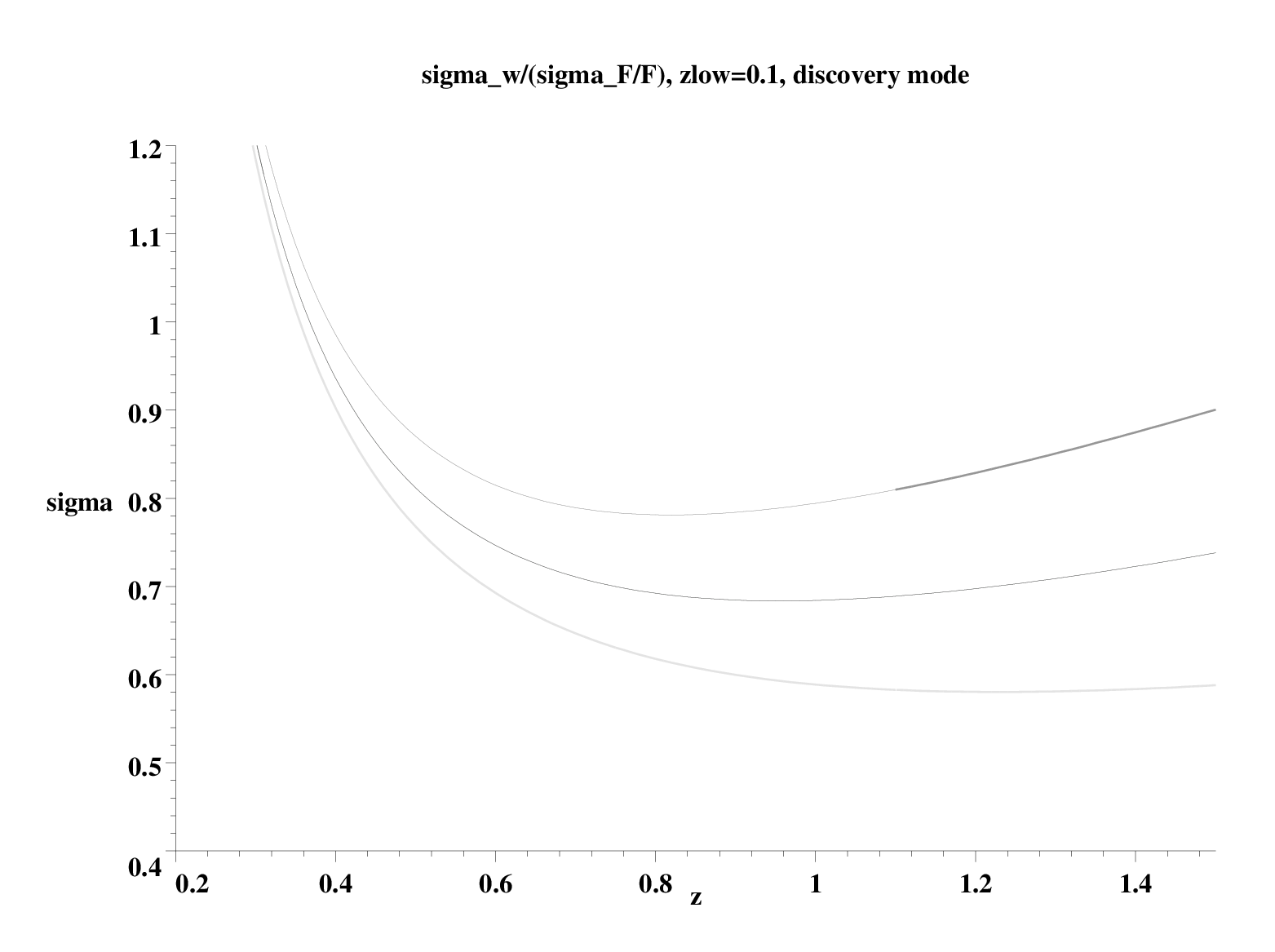,height=3in,width=3in}
\psfig{figure=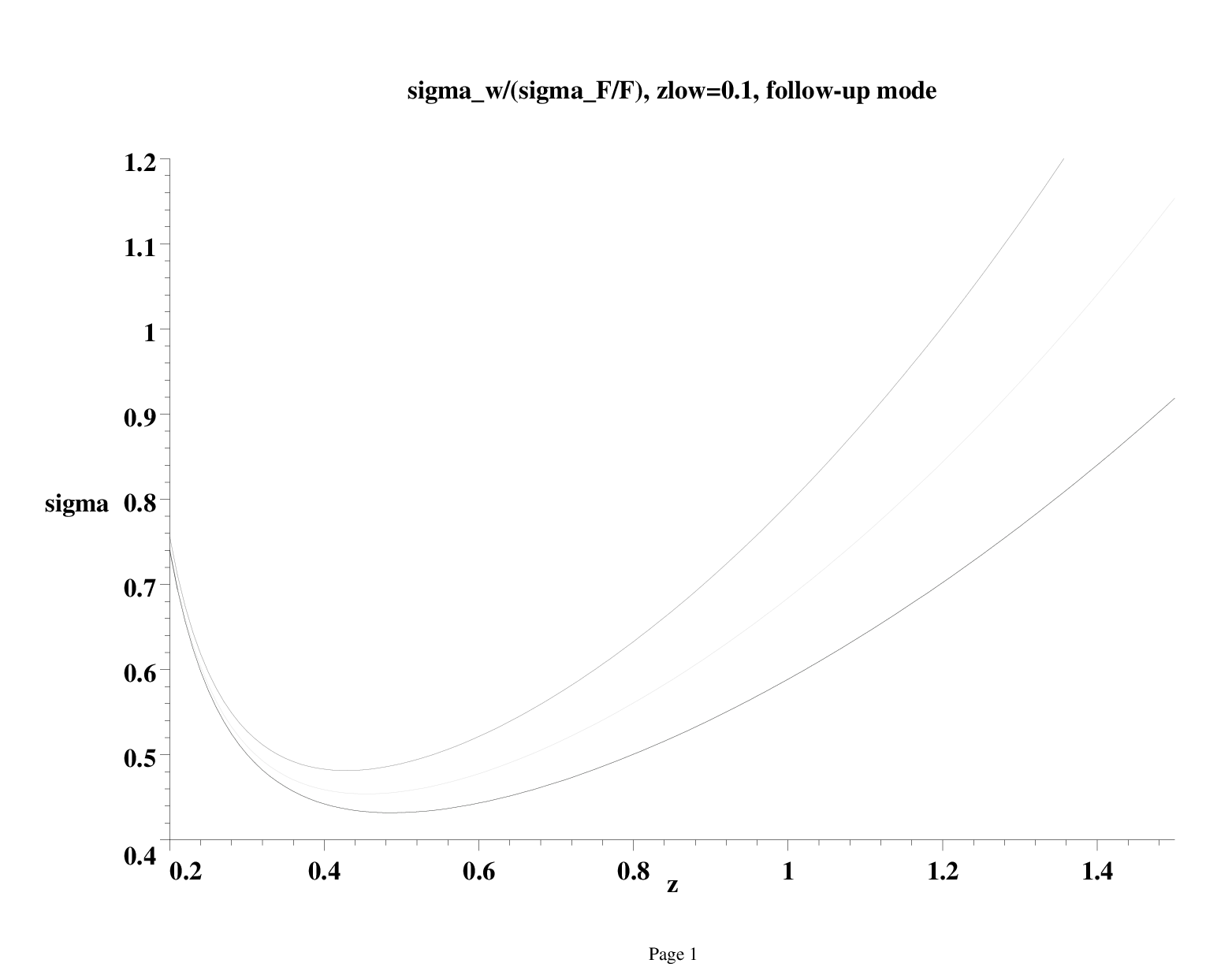,height=3in,width=3in} }
\caption{The statistical error, $\sigma$,  in the determination of $w$ from a survey 
of fixed duration conducted in a narrow range of redshifts around some redshift z,
  plotted as a function of  that redshift. The plots are in arbitrary units for $w=-1$, 
and $z_{low}=0.1$.  
The left hand plot is for \lq\lq{discovery mode}\rq\rq ($N\propto (1+z)^{-2}$);
the right hand plot is for \lq\lq{follow up mode}\rq\rq ($N\propto (1+z)^{-6}$).
In each, curves are for $w=-1.3$ (top), $w=-1$ (middle) and $w=-0.6$ (bottom).
}
\end{figure}

\hfil\break
\noindent
GDS would like to acknowledge fruitful discussions with J. Freeman, D. Huterer, E. Linder, S. Perlmutter,
and  M. Turner. Particular thanks to M. Turner for pointing out a sign error in an earlier version
as well as other useful comments.


\begin{thebibliography}{99}

\bibitem{SNCP}
S. Perlmutter {\it et al},  Astrophys.J. {\bf 517}, 565 (1999).

\bibitem{HRSS}
A.G. Riess {\it et. al.}  Astron. J. {\bf 116},  1009-1038 (1998);
P. M. Garnavich {\it et al.},  Ap. J. {\bf 493},  L53-57k (1998).

\bibitem{Boomerang}
C.B. Netterfield, P.A.R. Ade, J.J. Bock, J.R. Bond, J. Borrill, A. Boscaleri, K. Coble, C.R. Contaldi, 
B.P. Crill, P. de Bernardis, P. Farese, K. Ganga, M. Giacometti, E. Hivon, V.V. Hristov, A. Iacoangeli, 
A.H. Jaffe, W.C. Jones, A.E. Lange, L. Martinis, S. Masi, P. Mason, P.D. Mauskopf, A. Melchiorri, T. Montroy, 
E. Pascale, F. Piacentini, D. Pogosyan, F. Pongetti, S. Prunet, G. Romeo, J.E. Ruhl, F. Scaramuzzi
astro-ph/0104460.

\bibitem{MAXIMA}
 R. Stompor , M. Abroe, P. Ade, A. Balbi, D. Barbosa, J. Bock, J. Borrill, A. Boscaleri, P. De Bernardis,
P.G. Ferreira, S. Hanany, V. Hristov, A.H. Jaffe, A.T. Lee, E. Pascale, B. Rabii, P.L. Richards, G.F. Smoot, 
C.D. Winant, J.H.P. Wu,
Astrophys.J. 561 (2001) L7-L10.

\bibitem{DASI}
E. M. Leitch, C. Pryke, N. W. Halverson, J. Kovac, G. Davidson, S. LaRoque, E. Schartman, J. Yamasaki,
J. E. Carlstrom, W. L. Holzapfel, M. Dragovan, J. K. Cartwright, B. S. Mason,
      S. Padin, T. J. Pearson, M. C. Shepherd, A. C. S. Readhead 
astro-ph/0104488;
N. W. Halverson, E. M. Leitch, C. Pryke, J. Kovac, J. E. Carlstrom, W. L. Holzapfel, M. Dragovan, 
J. K. Cartwright, B. S. Mason, S. Padin, T. J. Pearson, M. C. Shepherd, A. C. S.  Readhead 
astro-ph/0104489;
C. Pryke, N. W. Halverson, E. M. Leitch, J. Kovac, J. E. Carlstrom, W. L. Holzapfel, M. Dragovan 
astro-ph/0104490.

\bibitem{parameters}
Recent reviews include: M. Kamionkowsky and A. Kosowsky, Ann. Rev. Nucl. Part. Sci. {\bf 49} 77 (1999) and W. Hu and S. Dodelson, Ann. Rev. Astron. Astrophys. {\bf in press} (2002).
For expectations from analysis of MAP data see, most recently
A. Kosowsky, M. Milosavljevic, R. Jimenez, astro-ph/0206014.

\bibitem{previous}
Previous work on the design of supernova surveys for cosmology includes most recently
D. Huterer, M. Turner, Phys. Rev. D, 64, 123527 (2001); also
M. Goliath, R. Amanullah, P. Astier, A. Goobar, R. Pain astro-ph/0104009;
E. Linder, astro-ph/0108280;
D.J. Eisenstein, W. Hu, M. Tegmark astro-ph/9805239;
W. Hu, D.J. Eisenstein, M. Tegmark, M. White  Phys.Rev. D59 (1999) 023512.


\end{thebibliography}
\end{document}